\documentclass[%
 reprint,
 superscriptaddress,
 amsmath,
 aps, prl,
]{revtex4-1}

\usepackage{graphicx}
\usepackage{dcolumn}
\usepackage{bm}
\usepackage{xcolor}

\definecolor{mygreen}{rgb}{0., 0.5, 0.0}

\usepackage{soul}

\begin{document}

\preprint{APS/123-QED}

\title{Electron-phonon coupling in the charge density wave state of CsV$_3$Sb$_5$}

\author{Yaofeng Xie}
\thanks{These authors made equal contributions to this work.}
\affiliation{Department of Physics and Astronomy, Rice University, Houston, Texas 77005, USA}
\author{Yongkai Li}
\thanks{These authors made equal contributions to this work.}
\affiliation{Centre for Quantum Physics, Key Laboratory of Advanced Optoelectronic Quantum Architecture and Measurement (MOE), School of Physics, Beijing Institute of Technology, Beijing 100081, China}
\affiliation{Beijing Key Lab of Nanophotonics and Ultrafine Optoelectronic Systems, Beijing Institute of Technology, Beijing 100081, China}
\affiliation{Material Science Center, Yangtze Delta Region Academy of Beijing Institute of Technology, Jiaxing, 314011, China}
\author{Philippe Bourges}
\email{philippe.bourges@cea.fr}
\affiliation{Laboratoire L$\acute{e}$on Brillouin, CEA-CNRS, Universit$\acute{e}$ Paris-Saclay, CEA Saclay, 91191 Gif-sur-Yvette, France}
\author{Alexandre Ivanov}
\affiliation{Institut Laue-Langevin, 71 avenue des Martyrs CS 20156, 38042 Grenoble Cedex 9, France}
\author{Zijin Ye}
\affiliation{Wuhan National High Magnetic Field Center \& School of Physics, Huazhong University of Science and Technology, Wuhan 430074, China}
\author{Jia-Xin Yin}
\author{M Zahid Hasan}
\affiliation{Laboratory for Topological Quantum Matter and Advanced Spectroscopy (B7), Department of Physics, Princeton University, Princeton, New Jersey 08544, USA}
\author{Aiyun Luo}
\affiliation{Wuhan National High Magnetic Field Center \& School of Physics, Huazhong University of Science and Technology, Wuhan 430074, China}
\author{Yugui Yao}
\affiliation{Centre for Quantum Physics, Key Laboratory of Advanced Optoelectronic Quantum Architecture and Measurement (MOE), School of Physics, Beijing Institute of Technology, Beijing 100081, China}
\affiliation{Beijing Key Lab of Nanophotonics and Ultrafine Optoelectronic Systems, Beijing Institute of Technology, Beijing 100081, China}
\author{Zhiwei Wang}
\email{zhiweiwang@bit.edu.cn}
\affiliation{Centre for Quantum Physics, Key Laboratory of Advanced Optoelectronic Quantum Architecture and Measurement (MOE), School of Physics, Beijing Institute of Technology, Beijing 100081, China}
\affiliation{Beijing Key Lab of Nanophotonics and Ultrafine Optoelectronic Systems, Beijing Institute of Technology, Beijing 100081, China}
\affiliation{Material Science Center, Yangtze Delta Region Academy of Beijing Institute of Technology, Jiaxing, 314011, China}
\author{Gang Xu}
\affiliation{Wuhan National High Magnetic Field Center \& School of Physics, Huazhong University of Science and Technology, Wuhan 430074, China}
\author{Pengcheng Dai}
 \email{pdai@rice.edu}
\affiliation{Department of Physics and Astronomy, Rice University, Houston, Texas 77005, USA}

\date{\today}

\begin{abstract}
Metallic materials with kagome lattice structure are interesting because their electronic structures 
can host flat bands, Dirac cones, and van Hove singularities, resulting in 
strong electron correlations, nontrivial band topology, charge density wave (CDW), and unconventional superconductivity. 
Recently, kagome lattice compounds $A$V$_3$Sb$_5$ ($A=$ K, Rb, Cs) are found to have intertwined CDW order and superconductivity. The origin of the CDW has been suggested to be purely electronic, arising 
from Fermi-surface instabilities of van Hove singularity (saddle point) near the $M$ points. Here we use neutron scattering experiments to demonstrate that
the CDW order in CsV$_3$Sb$_5$ is associated with static lattice distortion and a sudden hardening of the $B_{3u}$ longitudinal optical phonon mode, thus establishing that electron-phonon coupling must also play an important role in the CDW order of $A$V$_3$Sb$_5$.
\end{abstract}

\maketitle

Two-dimensional (2D) correlated transition metal materials with nearly square lattice structures have been heavily investigated because they display exotic properties such as
 unconventional superconductivity, electronic nematic phase, and intertwined charge, spin, and lattice order \cite{Fradkin2015,Dai2015}.  The 2D kagome lattice metallic materials, where atoms are arranged
into layered sets of corner-sharing triangles \cite{syozi}, are interesting because their electronic structures 
can host flat bands with quenched kinetic energy \cite{Sutherland,Leykam,Mazin2014,JXYin2019}, Dirac cones \cite{Ye2018,JXYin2018}, and van Hove singularities, resulting in 
strong electron correlations, nontrivial band topology, charge density wave (CDW) \cite{SLYu2012,WSWang2013,Kiesel2013}, 
and unconventional superconductivity \cite{Ortiz2019,Ortiz2020}. Recently, CDW order and superconductivity have been discovered to coexist in kaogme lattice metals $A$V$_3$Sb$_5$ ($A=$ K, Rb, Cs) [Figs. 1(a,b)] \cite{Ortiz2020,YXJiang2021,Liang2021}. In general, CDW order may originate from Fermi-surface instability following the 
Peierls’ description of an electronic instability in a one-dimensional chain of
atoms \cite{Peierls,WKohn,Rice1975,Johannes2006} or strong electron phonon coupling (EPC)/electron-electron correlations \cite{Zaanen1989,varma,XZhu2015}. Using inelastic X-ray scattering and angle resolved photoemission spectroscopy (ARPES), it was found that the CDW here may have purely electronic origin arising 
from Fermi-surface instabilities of van Hove singularity (saddle point) near the $M$ points 
without the involvement of lattice and electron-phonon coupling [Figs. 1(c) and 2(a)] \cite{HaoxiangLi2021}. If this is indeed the case \cite{Park2021,Kang2021}, the CDW order and superconductivity may 
intertwine in $A$V$_3$Sb$_5$ to form the exotic roton pair-density wave superconductivity and Majorana zero mode \cite{Liang2021,HChen2021,HZhao2021}.  Therefore, to understand the
electron pairing mechanism of superconductivity in $A$V$_3$Sb$_5$, one must first unveil the microscopic origin of the 
CDW order.

A CDW order resulting from a pure electronic origin via Fermi surface nesting should not distort the crystalline lattice or exhibit phonon anomaly, as 
suggested from inelastic X-ray scattering experiments \cite{HaoxiangLi2021}.  However, recent ARPES \cite{HLLuo2021} and optical spectroscopy \cite{Uykur2021} measurements in $A$V$_3$Sb$_5$ 
suggest that the momentum dependent EPC plays an important role in inducing the CDW transition.  Since neutrons  
cannot detect translational symmetry-breaking 
electron charge distribution but are sensitive to lattice distortion and phonon anomaly 
induced by the CDW order, we use neutron scattering to confirm the charge order and search for phonon anomaly across the CDW order
temperature.

In this paper, we report elastic and inelastic neutron scattering studies of CsV$_3$Sb$_5$, which exhibits CDW order at $T_{CDW}=94$ K and superconductivity
at $T_c=2.5$ K \cite{Ortiz2020}. We observe in-plane 2 by 2 superlattice peak below $T_{CDW}=94$ K, thus revealing that the CDW order in CsV$_3$Sb$_5$ is  associated with
lattice distortion and not a pure electronic instability \cite{Johannes2006,XZhu2015}. However, we find no detectable changes in charge ordering intensity across $T_c$. Furthermore, we use inelastic neutron scattering
to map out longitudinal acoustic and optical phonon modes near Bragg peak position $(3,0,0)$ at temperatures 
across $T_{CDW}$ and $T_c$ (Figs. 1-4). 
While acoustic phonon mode show no dramatic change across $T_{CDW}$ 
consistent with the earlier work by inelastic X-ray scattering \cite{HaoxiangLi2021}, we find that 
optical phonon mode with $B_{3u}$ symmetry [Fig. 1(d,f)], possibly associated with the 2 by 2 charge order and 
inverse Star of David deformation of the kagome lattice
 \cite{Tan2021,BROrtiz2021,Ratcliff2021,HMiao2021}, hardens across $T_{CDW}$ at the $M$ point.  
We also identified an optical phonon mode near the expected energy for
 the phonon mode with $B_{1u}$ symmetry at $\Gamma$ point [Fig. 1(d)], and found it to have 
no observable changes
across  $T_{CDW}$. Therefore, our results firmly establish that the CDW order in CsV$_3$Sb$_5$ is
not a purely electronic transition, and the EPC must also play an important role in the formation of CDW order.

\begin{figure}[t]
\centering
\includegraphics[scale=.25]{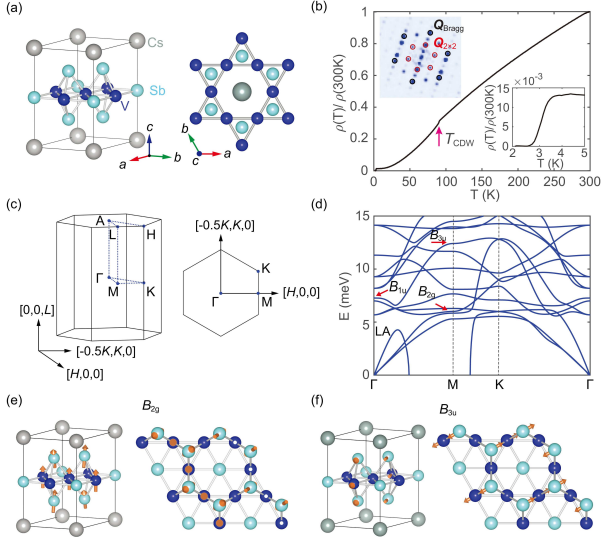}
\caption{(a) Crystal structure of ${\rm CsV_3Sb_5}$ from three-dimensional view (left) and top view (right). (b) Temperature dependence of 
resistivity of ${\rm CsV_3Sb_5}$, where CDW order is seen around 95 K marked by the arrow. The upper inset is the Fourier transform of Sb topographic image from STM measurement, showing the ordering peaks (${\rm Q_{2\times2}}$) and Bragg peaks (${\rm Q_{Bragg}}$) \cite{YXJiang2021}. 
The lower inset shows superconducting transition temperature of the sample.
(c) 3D and 2D Brillouin zone of ${\rm CsV_3Sb_5}$. The high symmetry points are specified. (d) DFT calculated phonon spectra of ${\rm CsV_3Sb_5}$. Full vanadium breathing vibration associated with the $B_{1u}$ mode is marked by the arrow around 7 meV [see Fig. 4(a)].
The $B_{2g}$ and $B_{3u}$ modes at the $M$ point are labeled. Lattice distortion for the $B_{2g}$ (e) and $B_{3u}$ (f) modes at the $M$ point in three-dimensional view (left) and top view (right).}
\end{figure}

\begin{figure}[t]
\centering
\includegraphics[scale=.3]{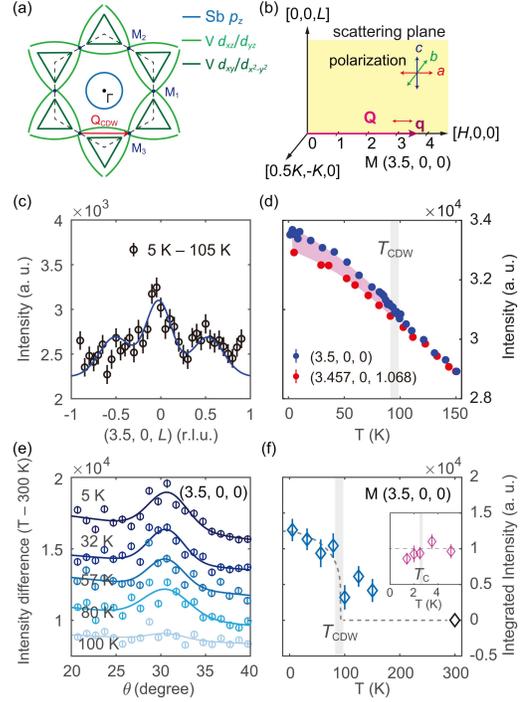}
\caption{(a) Schematics of the Fermi surface of ${\rm CsV_3Sb_5}$, where the circular Fermi surface near $\Gamma$ is from
Sb $p_z$ band, and vanadium 
$d_{xz}/d_{yz}$ and $d_{xy}/d_{x^2-y^2}$ bands are shown in light and dark green, respectively \cite{Ortiz2020,Kang2021,HLLuo2021}.  
The CDW order is suggested to be associated with Fermi surface instability of the $M$ points (red arrow).
(b) Schematics of the $[H,0,L]$ scattering plane. The double-headed arrows represent the phonon polarization directions. 
For wave vectors probed in the present experiment, we are mostly probing longitudinal phonon modes along the $\Gamma$-$M$ ($H$) direction.
(c) The temperature difference elastic scattering along the $[3.5,0,L]$ direction.  The raw data is shown in the supplementary information. The solid line is the Gaussian fit to the data. (d) Temperature dependence of the 
scattering at $(3.5,0,0)$ and $(3.475,0,1.068)$ positions \cite{SI}.
(e) Temperature differences of rocking curve 
scans across ${\bf Q} = (3.5,0,0)$ using 300 K data as background. The solid lines are Gaussian fits to the data. (f) Temperature dependence of the integrated intensity at $(3.5,0,0)$ across the $T_{CDW}$.  The inset shows temperature
dependence of the CDW order integrated intensity across $T_{c}$.}
\end{figure}

Our neutron scattering experiments were carried out at the IN8 thermal triple-axis spectrometer at the Institut Laue-Langevin (ILL), Grenoble, France. 
We used doubly focused pyrolytic graphite monochromator and analyzer with PG(0,0,2) reflection and fixed scattered (final) energy $E_f = 14.68$ meV. 
Several scans have been performed with a 2D-focusing Si(1,1,1) monochromator.
 Using a hexagonal lattice with $a = b = 5.495$ \AA, $c=9.309$ \AA\ 
as shown in Fig. 1(a) to describe its crystal structure, the momentum transfer $\textbf{Q}=H\textbf{a}^\ast+K\textbf{b}^\ast+L\textbf{c}^\ast$ is denoted as $(H,K,L)$ in reciprocal lattice units (r.l.u.) [Fig. 1(c)] \cite{Ortiz2020}. About four hundred individual single crystals were co-aligned on four aluminum plates to form an assembly with a volume of 0.11 ${\rm cm^3}$ and an in-plane mosaic 
spread of 3.5 degrees \cite{SI}. The crystal assembly was put inside a He cryostat and oriented in the $[H,0,L]$ horizontal scattering plane. We also use density functional theory (DFT) to calculate the phonon spectra similarly to previous work \cite{Tan2021}.

Figure 1(b) shows transport data for CsV$_3$Sb$_5$, confirming the existence of  CDW order below $T_{CDW}=95$ K and
superconductivity below $T_c\approx 2.5$ K.  Previous scanning tunneling microscopy results showing 2 by 2 charge ordering is shown in the upper inset \cite{YXJiang2021}.
Figures 1(d,e,f) summarize DFT calculated phonon spectra 
at ambient pressure and the symmetries of the most interesting modes near the $M$ point in reciprocal
space. Consistent with the previous work \cite{Tan2021}, we find that the longitudinal acoustic phonon mode is unstable at ambient pressure, 
suggesting that this mode may be relevant to the formation of CDW order. At the Brillouin zone boundary $M$ point, we expect to observe 
optical phonon modes with $B_{2g}$ and $B_{3u}$ symmetry, corresponding to out of plane and half breathing mode of vanadium as shown in Figs. 1(e) and 1(f), respectively.

Figure 2 summarizes the key results from our neutron diffraction experiments to probe the temperature dependence of the lattice distortion induced by
the CDW order.  From ARPES experiments, it was found that the electronic structure of 
of $A$V$_3$Sb$_5$ is dominated by vanadium bands near $M$ points [Fig. 2(a)] \cite{Ortiz2020,HLLuo2021,ZGWang2021}. 
Therefore, the 2 by 2 CDW order may arise from Fermi surface nesting of qusiparticle excitations between three $M$ points. 
Since our crystal assembly is aligned in the $[H,0,L]$ scattering zone [Fig. 2(b)], we can probe elastic scattering as well as phonons around nuclear
Bragg peak $(3,0,0)$ ($\Gamma$) position. Figure 2(c) shows temperature difference plot of the $[3.5,0,L]$ elastic scan between 5 K and 100 K.  We find a clear peak centered at $L=0$ and 
weaker peaks at $L=\pm 0.5$, thus establishing the presence of 
low-temperature lattice distortion in CsV$_3$Sb$_5$. 
To confirm that the temperature dependent lattice distortion is associated 
with CDW order \cite{Ortiz2020,YXJiang2021,Liang2021,HaoxiangLi2021}, we measure temperature
dependence of the scattering at the $(3.5,0,0)$ (signal) 
and $(3.457,0,1.068)$ (background) positions, revealing
clear intensity gain of the $(3.5,0,0)$ scattering approximately
below $T_{CDW}$ [Fig. 2(d)].  Figure 2(e) shows rocking curve scans around $(3.5,0,0)$ at temperatures across $T_{CDW}$ using 300 K 
data as background scattering.  
While the scattering is featureless at 100 K, a clear peak centered at $(3.5,0,0)$ appears at temperatures
below $T_{CDW}$ [Figure 2(e)].  Figure 2(f) shows temperature dependence of the integrated intensity, again confirming
the appearance of the CDW peak below $T_{CDW}$.  However, the intensity of CDW peak does not seem to change across 
the superconducting transition temperature 
$T_c$ [see inset of Fig. 2(f)].

Figures 3(a) and 3(b) show constant-${\bf Q}$ scans at ${\bf Q}=(2.5,0,0)$ (the 
$M$ point) and $(2.7,0,0)$ (approximate middle of the Brillouin zone), respectively, 
from room temperature to 5 K across $T_{CDW}$.  
These scans show two weakly dispersive phonon modes at $E\approx 6$ meV and  $\sim$10 meV. 
While the $\sim$10 meV mode at the $M$ point shows a clear $\sim$2 meV  hardening below $T_{CDW}$,
the $\sim$6 meV mode only hardens slightly on cooling, and 
has negligible changes across $T_{CDW}$ [Fig. 3(c)].  Similar behavior is seen in the full-width-half-maximum (FWHM) of the 
$\sim$10 meV mode, but not the  $\sim$6 meV mode [Fig. 3(e)]
For comparison, these two phonon modes at ${\bf Q}=(2.7,0,0)$ show no observable anomaly across $T_{CDW}$ in energy 
position [Fig. 3(d)] and width [Fig. 3(f)].  The wave vector dependence of the FWHM of the $\sim$6 meV and $\sim$10 meV modes
at temperatures above (100 K and 300 K) and below (5 K)  $T_{CDW}$ is shown in Figs. 3(g) and 3(h). Consistent with
Figs. 3(a-f), the $\sim$10 meV mode shows clear broadening below $T_{CDW}$ at the $M$ point.

\begin{figure}[t]
\centering
\includegraphics[scale=.3]{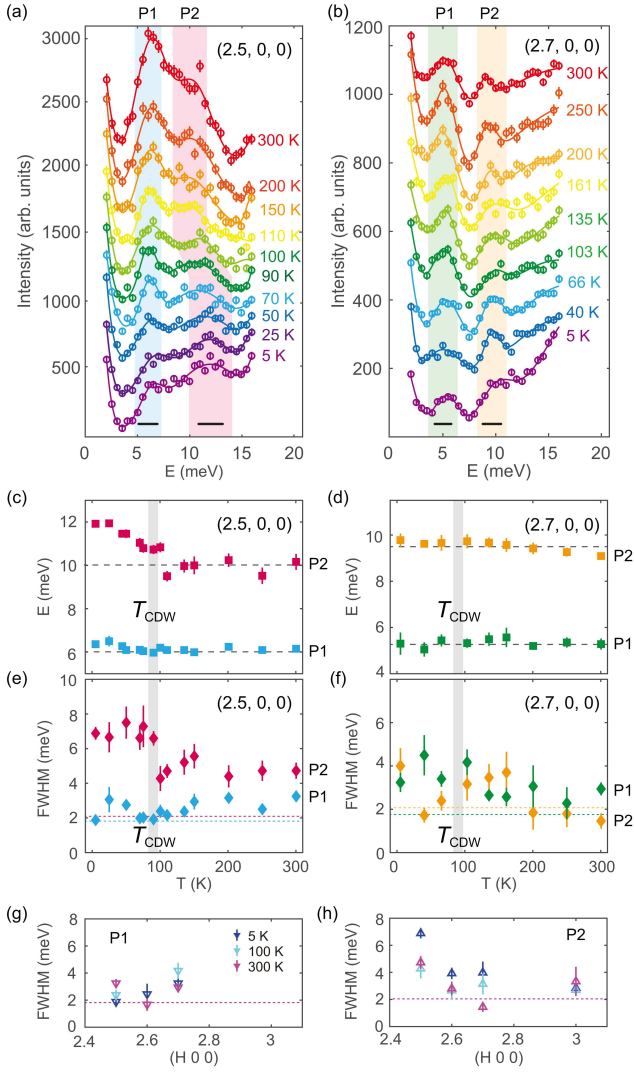}
\caption{(a,b) Temperature dependence of the constant-${\bf Q}$ scans at ${\bf Q}=(2.5,0,0)$ and $(2.7,0,0)$, respectively. The color-shaded regions highlight the 
two peaks, P1 and P2, observed by inelastic neutron scattering. Solid lines are results from multiple Gaussian fits. The scans are shifted vertically for clarity. The horizontal bars indicate instrumental energy resolution at different energies.  
(c,d) Temperature dependence of the phonon energy at $(2.5 0 0)$ and $(2.7 0 0)$, respectively. The phonon mode at 
${\bf Q}=(2.5,0,0)$ hardens below $T_{CDW}$. (e,f) Temperature dependent phonon energy line-widths of the 
P1 and P2 mode at $(2.5,0,0)$ and $(2.7,0,0)$, respectively. (g,h) Wave vector dependence of the phonon line-widths of the P1 and P2 modes, respectively. 
The shaded vertical bars in (c-f) mark the temperature of CDW order in CsV$_3$Sb$_5$ and horizontal dashed lines are guides to the eye.
}
\end{figure}

\begin{figure}[t]
\centering
\includegraphics[scale=.25]{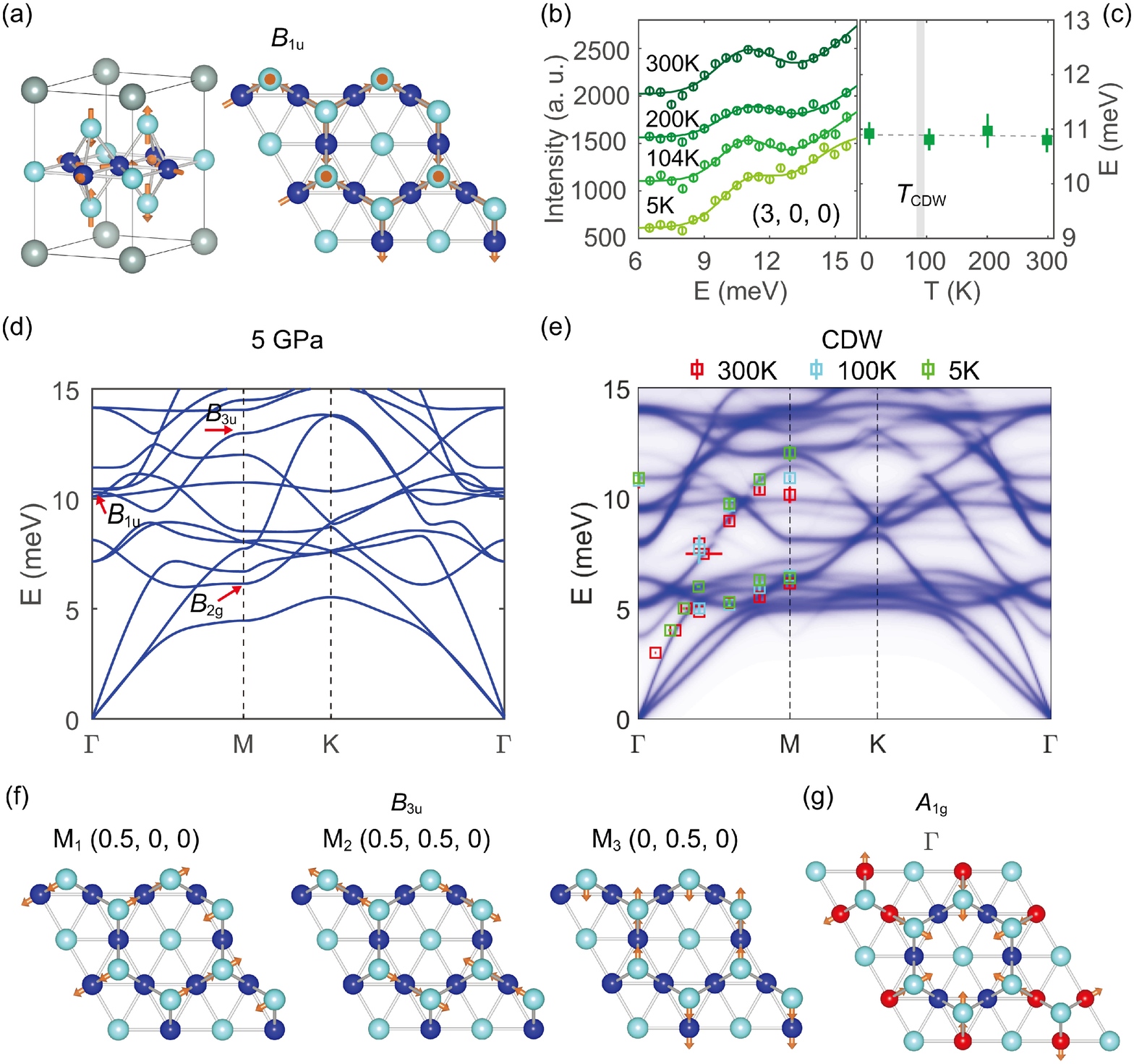}
\caption{(a) Lattice distortion of the full breathing mode with the $B_{1u}$ symmetry at the $\Gamma$ point in 
three-dimensional view (left) and top view (right). (b) Temperature dependence of the constant-${\bf Q}$ scans at $(3,0,0)$, where a phonon mode 
at 11 meV is seen. The solid lines are Gaussian fits to the data. The scans are shifted vertically for clarity. 
(c) Temperature dependence of the phonon mode energy at $(3,0,0)$. (d) DFT calculated phonon spectra at a hydrostatic pressure of 5 GPa.  The $B_{1u}$ mode at the $\Gamma$ point is labeled. We note that 
 the  $B_{1u}$ mode energy is shifted from 7 meV at 0 GPa to 11 meV at 5 GPa.
 (e) Comparison of DFT calculated phonon spectra in the CDW phase with inelastic neutron scattering determined phonon dispersions  
measured at 5, 100, and 300 K. 
The calculation is unfolded to the non-CDW phase to compare with the neutron data. (f) Vadanium vibrational mode with $B_{3u}$ symmetry
at three $M$ positions. (g) Sum of the three $B_{3u}$ mode has $A_{1g}$ symmetry
at $\Gamma$ point in the CDW state. }
\end{figure}

To understand the microscopic origin of the observed phonon spectra, we compare the data in Fig. 3 with the phonon spectra 
calculated from DFT [Figs. 4(a,d,e)].  Figure 4(b) shows constant-${\bf Q}$ scans
at ${\bf Q}=(3,0,0)$ zone center below and above $T_{CDW}$, revealing a clear phonon mode at $E\approx 11$ meV. 
Temperature independence of the mode energy 
between 5 K and 300 K shown in Fig. 4(c) suggests that the mode is not obviously affected by the CDW order at 95 K. 
Therefore, we conclude that this phonon mode is irrelevant to the CDW order. 
Comparing energy of the observed mode with DFT calculation under 5 GPa [Fig. 4(d)], we conclude that the 11 meV mode is likely to be the optical phonon mode
with $B_{1u}$ symmetry [Figs. 4(a)]. Figure 4(e) compares the measured and calculated dispersion curves for CsV$_3$Sb$_5$. 
Inspection of the Figure reveals that the measured longitudinal acoustic phonon mode agrees with the DFT calculation unfolded in the 
CDW state reasonably well. This means that the energy of the zone boundary longitudinal acoustic phonon mode occurs at approximately
6 meV.  Since temperature dependence of the mode energy shows a slight hardening below $T_{CDW}$ [Fig. 3(c)], our results 
indicate no softening of the acoustic phonon modes at the $M$ point in the CDW state, consistent
with the inelastic X-ray scattering work \cite{HaoxiangLi2021}. On the other hand, the 
optical phonon mode at $\sim$10 meV is likely associated with half breathing mode of vanadium with
 $B_{3u}$ symmetry [Fig. 4(e)].

As discussed in Ref. \cite{XZhu2015}, the classical picture of Fermi surface nesting induced CDW order from 
Peierls’ description \cite{Peierls,WKohn} fails in many real systems and the wave vector dependence of the EPC matrix element determines
the characteristic of the CDW phase.  For example, CDW order in NbSe$_2$ is not due to Fermi surface nesting but instead arises 
from the EPC as it is seen phonon energy softening, broadened phonon line-width at the CDW ordering wave vector  \cite{XZhu2015}.
The situation in CsV$_3$Sb$_5$ is somewhat different. While there is no evidence of acoustic phonon softening and broadening at the
charge ordering wave vector (Figs. 3 and 4) consistent with the weakly first order
nature of the CDW transition \cite{Ortiz2020,HaoxiangLi2021}, the energy of 
optical phonon mode with $B_{3u}$ symmetry shows a clear hardening below $T_{CDW}$ [Fig. 3(c)].  
In addition, the optical phonon line-width broadens  
at the CDW wave vector below $T_{CDW}$ [Fig. 3(h)].  Since charge order occurs at three equivalent $M$ points, superposition of three $B_{3u}$ modes by band folding in the CDW state
can lead to an inverse Star of
David deformation of a vanadium breathing mode with $A_{1g}$ symmetry and two other degenerate modes that do not
have three fold rotational symmetry [Figs. 4(f,g)]. Therefore, our results provide strong evidence that the EPC must play an important role in the
formation of the CDW order in CsV$_3$Sb$_5$. Although recent $\mu$SR measurements suggest that CDW order is also associated with a time reversal symmetry breaking field \cite{Mielke}, consistent with the presence of a chiral flux phase in the CDW state of CsV$_3$Sb$_5$ \cite{Feng2021,Lin2021}, it is unclear how the time reversal symmetry breaking field in the CDW state induced by flux phase can affect the lattice and its vibrations.

In summary, we have carried out elastic and inelastic neutron scattering experiments on CsV$_3$Sb$_5$.  Our elastic results confirm the presence of 2 by 2 charge order below 95 K, indicating that the CDW order also involves lattice distortion. By comparing phonons measured by inelastic neutron scattering experiments with that of DFT calculations,
we conclude that acoustic phonons in CsV$_3$Sb$_5$ do not respond to CDW order but optical phonon mode with
$B_{3u}$ symmetry hardens below $T_{CDW}$ at $M$ points. This phonon hardening is likely associated with 
 an inverse Star of
David deformation of a vanadium atoms with $A_{1g}$ symmetry \cite{Tan2021}.  These results therefore indicate
that the effect of lattice must be taken into account to achieve a comprehensive understanding of the CDW state in
$A$V$_3$Sb$_5$.

P.D. is grateful to Binghai Yan, B. R. Ortiz, and Stephen Wilson for helpful discussions. 
 The neutron scattering and basic materials characterization work at Rice is supported
by the U.S. Department
of Energy, BES under Grant No. DE-SC0012311 and by the Robert A. Welch
Foundation under Grant No. C-1839, respectively (P.D.).
The work at Beijing Institute of Technology was supported by the National Key R\&D Program of China (Grant No. 2020YFA0308800), the Natural Science Foundation of China (Grant Numbers. 92065109, 11734003), the Beijing Natural Science Foundation (Grant No. Z190006, Z210006), and the Beijing Institute of Technology (BIT) Research Fund Program for Young Scholars (Grant No. 3180012222011). Z.W. thanks the Analysis \& Testing Center at BIT for assistance in facility support. Work at Huazhong University of Science and Technology was supported by the National Key Research and Development Program of China (2018YFA0307000), and the National Natural Science Foundation of China (11874022). Work at Princeton University was supported by the U.S. Department of Energy under Grant No. DOE/BES DEFG-02-05ER46200. 

{}

\end{document}